\documentclass[11pt,preprint]{aastex}
\newcommand{\CS}{CS~31062--050}
\newcommand{\LP}{LP~625--44}
\shorttitle{Ir and Os abundances in carbon-enhanced metal-poor stars}
\shortauthors{Aoki et al.}

\begin{document}
\title{Carbon-Enhanced Metal-Poor Stars. Osmium and Iridium Abundances in the Neutron-Capture-Enhanced Subgiants CS~31062--050 and LP~625--44\footnote{Based on
data collected at the Subaru Telescope, which is operated by the
National Astronomical Observatory of Japan.}}

\author{Wako Aoki\altaffilmark{2},
Sara Bisterzo\altaffilmark{3,4},
Roberto Gallino\altaffilmark{3,5}, 
Timothy C. Beers\altaffilmark{6},
John E. Norris\altaffilmark{7}, 
Sean G. Ryan\altaffilmark{8,9}, 
Stelios Tsangarides\altaffilmark{8}}

\altaffiltext{2}{National Astronomical Observatory, Mitaka, Tokyo,
181-8588, Japan; email: aoki.wako@nao.ac.jp}
\altaffiltext{3}{Dipartimento di Fisica Generale dell'Universita' di
Torino, via P. Giuria 1, 10125 Torino, Italy; bisterzo@ph.unito.it, gallino@ph.unito.it}
\altaffiltext{4}{Forschungszentrum Karlsruhe, Institute fuer Kernphysik, P.O. Box 3640, D-76021 Karlsruhe, Germany}
\altaffiltext{5}{Centre for Stellar and Planetary Sciences, School of
    Mathematical Sciences, Monash University, Building 28, 
    Victoria 3800, Australia}
\altaffiltext{6}{Dept. of Physics \& Astronomy, CSCE: Center for the Study of
Cosmic Evolution, and JINA: Joint Institute for Nuclear Astrophysics,
Michigan State University, E. Lansing, MI
48824; beers@pa.msu.edu}
\altaffiltext{7}{Research School of Astronomy and Astrophysics, The
Australian National University, Mount Stromlo Observatory, Cotter
Road, Weston, ACT 2611, Australia; email: jen@mso.anu.edu.au}
\altaffiltext{8}{Department of Physics and Astronomy, The Open
University, Walton Hall, Milton Keynes, MK7 6AA, UK; email:
stsangarides@gmail.com}
\altaffiltext{9}{Present address: Centre for Astrophysics Research, STRI and School of Physics,
Astronomy and Mathematics, University of Hertfordshire, College Lane,
Hatfield AL10 9AB, United Kingdom; s.g.ryan@herts.ac.uk} 

\begin{abstract} 

We have investigated the abundances of heavy neutron-capture elements,
including osmium (Os) and iridium (Ir), in the two Carbon-Enhanced
Metal-Poor (CEMP) subgiants {\CS} and {\LP}. {\CS} is known to be a
so-called CEMP-r/s star, which exhibits large excesses of s-process
elements such as barium (Ba) and lead (Pb), as well as a significant
enhancement of europium (Eu) that cannot be explained by conventional
s-process production in Asymptotic Giant Branch star models. Our
analysis of the high-resolution spectrum for this object has
determined, for the first time, the abundances of Ir and Os, elements
in the third peak of the r-process nucleosynthesis. They also exhibit
significant excesses relative to the predictions of standard s-process
calculations. These two elements are not detected in a similar-quality
spectrum of {\LP}; the derived upper limits on their abundances are
lower than the abundances in {\CS}. We compare the observed abundance
patterns of neutron-capture elements, including Os and Ir, in these
two stars with recent model calculations of the s-process, and discuss
possible interpretations.

\end{abstract} 


\section{Introduction}\label{sec:intro}

Recent abundance analyses of Carbon-Enhanced Metal-Poor (CEMP) stars
have provided a unique opportunity to investigate the abundance
patterns produced by s-process nucleosynthesis at low metallicity
\citep[e.g., ][]{norris97,hill00}.  Of particular importance, in the
past several years abundance measurements of neutron-capture elements
have been extended to cover a wide range of atomic numbers, including
heavy species such as lead \citep[Pb; e.g. ][]{aoki00, vaneck01} and
bismuth \citep[Bi; e.g. ][]{ivans05}. One interesting class of objects
that have recently been identified are the so-called r/s stars
\citep[e.g. ][]{hill00, cohen03, beers05, jonsell06}. These stars
exhibit large enhancements of the neutron-capture elements, whose
overall abundance patterns can be explained by the s-process, but
whose abundance of the r-process element Eu is much higher than the
value predicted by conventional s-process nucleosynthesis models
\footnote{The solar-system isotopic abundances of neutron-capture
elements are decomposed using s-process models. The fraction of
r-process contribution is estimated by subtracting the s-process
contribution from the total abundances. The terms s-process elements
and r-process elements mean the elements whose solar-system abundances
are mostly explained by corresponding processes. For instance, more
than 94\% of Eu in the Solar System is yielded by the r-process,
according to \citet{arlandini99} and \citet{simmerer04}.}. All of the
presently recognized r/s stars also exhibit large carbon enrichment
([C/Fe] $\gtrsim + 1.0$); \citet{beers05} refer to these as CEMP-r/s
stars.

%(I THINK THIS IS TRUE, BUT IT SHOULD BE CHECKED)

The observed excesses of s-process elements in CEMP stars are usually
explained by the yields of Asymptotic Giant Branch (AGB) stars (see
Herwig 2005 for a recent review). Since most CEMP stars observed today
are not in their AGB phase, it is usually assumed that the AGB star
responsible for the s-process (and carbon enhancement) was once the
primary of a binary system, and the elements formed by it were
transferred by a stellar wind to a lower-mass companion (the star now
observed) prior to the primary evolving to become a faint white
dwarf. One possible scenario to explain CEMP-r/s stars is to assume
that the parent cloud of the binary system was already enriched in the
r-process elements. The r-process nucleosynthesis source(s),
presumably core-collapse supernovae, would leave an elemental imprint
of its products on such very low-metallicity objects, according to the
above interpretation. One important observational constraint is that a
significant fraction of very metal-poor s-process-enhanced stars are
in fact r/s stars. For instance, \citet{jonsell06} list 17
r/s stars among 24 s-process-enhanced objects. This stands in contrast
to the fact that r-process-enhanced ([Eu/Fe]$\gtrsim +1.0$) stars with
no excess of s-process elements are quite rare among very metal-poor
stars ($\sim 5$\%; Barklem et al. 2005). The large fraction of r/s
stars among s-process-enhanced stars is not explained by assuming that
these stars accidentally had large overabundances of r-process
elements, but suggests that a close relationship may exist between the
s- and r-processes that form the abundance patterns of r/s stars.

%However, this assumption may raise important constraints on the origin
%of the r-process. 

%However, this assumption results in an important constraint on the
%origin of the r-process: since the excesses of s-process elements are
%usually explained by the yields of asymptotic giant branch (AGB)
%stars, the r-process nucleosynthesis should be closely related to such
%objects according to the above interpretation.

The interest in r/s stars has been rapidly growing in recent
years. However, the classification of such objects mostly relies on
the relatively easily measured Eu abundance, although the abundances
of elements between Gd and Lu also provide constraints on the origin
of neutron-capture elements for a small number of stars (e.g.,
CS~29497-030; Ivans et al. 2005). In order to better estimate the
relative contributions of the s- and r-processes in r/s stars, it is
important to increase the number of such stars with available
abundance measurements for elements in the third peak of the r-process
nucleosynthesis (Os and Ir, whose r-process fractions in solar-system
material are 91.6 and 98.8\%, respectively, according to
\citet{simmerer04}).

We have investigated high-resolution ultraviolet-blue spectra of two
CEMP subgiants, {\CS} and {\LP}, whose abundance patterns were studied
relatively well by previous work
\citep{norris97,aoki00,aoki02a,aoki02b,johnson04}. These stars have
quite similar atmospheric parameters and carbon enhancements.
Moreover, variations of radial velocities have been detected in both
of these stars, indicating that they both belong to binary systems. In
this Letter, the abundances of Os and Ir in {\CS}, as well as upper
limits on these elements in {\LP}, are reported. The abundance
patterns of these stars are then compared to the predictions of modern
s-process nucleosynthesis models.

\section{Observations and Analyses}

High-resolution spectra (3100--4700~{\AA}) of {\CS} and {\LP} were
obtained in August 2002 with the High Dispersion Spectrograph \citep[HDS:
][]{noguchi02} of the Subaru Telescope. The spectra
were used to measure the Eu isotope ratios in these stars in our
previous work \citep{aoki03}: see \S 2 of that paper for details of
the observations and data reduction. The photon counts (per
0.9~km~s$^{-1}$ pixel) at 3515~{\AA} and 4260~{\AA}, where measurable
Ir and Os lines exist, are 8300 and 25000 for {\CS} and 6800 and 20500
for {\LP}, respectively. Figure \ref{fig:iros} shows the spectra of
these stars in the region of these spectral lines. The high resolving
power ($R=90,000$) achieved in these observations is very important to
resolve the Ir line from the other absorption features. In this paper,
we also use the red spectrum of {\CS} obtained by \citet{aoki06}
to determine the Ba abundance.

We carry out LTE abundance analyses for {\CS} and {\LP}, using the
model atmospheres of \citet{kurucz93}, and adopting the atmospheric
parameters determined by \citet{johnson04} and \citet{aoki02a},
respectively (Table~\ref{tab:res}).
One reason for adopting these
values from previous work is to combine our abundances of Os and Ir
with previous results for other elements. Note that the effective
temperatures estimated from the $V-K$ colors of these two stars,
adopting the scale of \citet{alonso99}, are 150--180~K higher than the
values adopted here (5500~K). However, this small difference does not
significantly affect the relative abundance patterns discussed in this
Letter.

%For the LTE abundance analyses using model atmospheres of
%\citet{kurucz93}, the atmospheric parameters of {\CS} and {\LP}
%determined by \citet{johnson04} and \citet{aoki02a}, respectively,
%were adopted 

%($T_{\rm eff}$/$\log g$/[Fe/H]/$v_{\rm turb}$ =
%5500~K/2.7/$-2.2$/1.3~km~s$^{-1}$ for {\CS} and
%5500~K/2.5/$-2.7$/1.2~km~s$^{-1}$ for {\LP}).

We applied the spectrum synthesis technique to the Ir
$\lambda$3513~{\AA} and Os $\lambda$4260~{\AA} lines, as illustrated
in Figure \ref{fig:iros}. The oscillator strengths of these lines are
taken from the list of \citet{hill02}.  We note that their oscillator
strength of the Os line is confirmed by the recent measurements of
\citet{quinet06}. As a test, we also carry out analyses of Ir and
Os, based on these same lines, for the spectra of CS~31082--001 (Hill
et al.  2002) and HD~6268 \citep{honda04}, and find good agreement
between our results and those of these authors (the differences are
smaller than 0.1~dex). The Ir and Os lines are clearly detected in the
spectrum of {\CS}, and abundances of these species are determined for
this star, while only upper limits on the abundances of these species
are estimated for {\LP}. The results are listed in Table
\ref{tab:res}.

%For both stars, we also measured the abundances of the six heavy
%neutron-capture elements Ba, La, Eu, Yb, Hf and Pb. 

We also measure abundances for six heavy neutron-capture elements, Ba,
La, Eu, Hf, and Pb, for both stars.  The Ba abundance of {\CS} is
re-determined using the $\lambda 5853$~{\AA} and $\lambda 6141$~{\AA}
lines, which are much weaker, and less sensitive to the effect of
hyperfine splitting than the resonance lines, and thus preferable for
abundance measurements. The Ba abundance we derive is 0.2~dex lower
than that reported by \citet{johnson04}, who obtained an extremely
large enhancement of this element compared to other neutron-capture
species. Our measurement, using different spectral lines, confirms the
large excess of Ba, although the enhancement is not as large as
reported by \citet{johnson04}. We confirm a good agreement between our
results and those of \citet{johnson04} for other elements in {\CS}. We
note that the Yb abundance is determined using the $\lambda
3476$~{\AA} line, rather than the very strong $\lambda 3692$~{\AA}
line, in order to avoid uncertainties due to damping effects and
isotope splitting. The same line is used to determine the Yb
abundance of {\LP}. \citet{aoki02a} analyzed the $\lambda 3692$~ {\AA}
line in this star, but only obtained a very uncertain result. The La
abundance of {\LP} is re-determined using the line data provided by
\citet{lawler01}. The errors given in Table~\ref{tab:res} are the
uncertainties due to fitting of synthetic spectra. Errors due to the
uncertainties of atmospheric parameters are of the order of
0.1--0.15~dex, as estimated by previous work for heavy neutron-capture
elements \citep[e.g., ][]{aoki02a}. It should be noted that no
stronger constraint on the upper limit of Th abundance in {\CS} than
that of \citet{johnson04} is obtained, because of the severe blending
of CH molecular features with the \ion{Th}{2} $\lambda$4019~{\AA} line,
although the spectral resolution of our data is higher than theirs.

In the following discussion we adopt the results of the present
analysis for Ba, Os, and Ir for {\CS}, and those of \citet{johnson04}
for other elements. For {\LP}, the abundance of La and Yb, as well as
the upper limits on the Os and Ir abundances, determined by the
present work, are combined with the results of \citet{aoki02a}.

\section{Comparison with Modern s-Process Models}

\begin{figure} 
\begin{center}
\includegraphics[angle=-90, width=10cm]{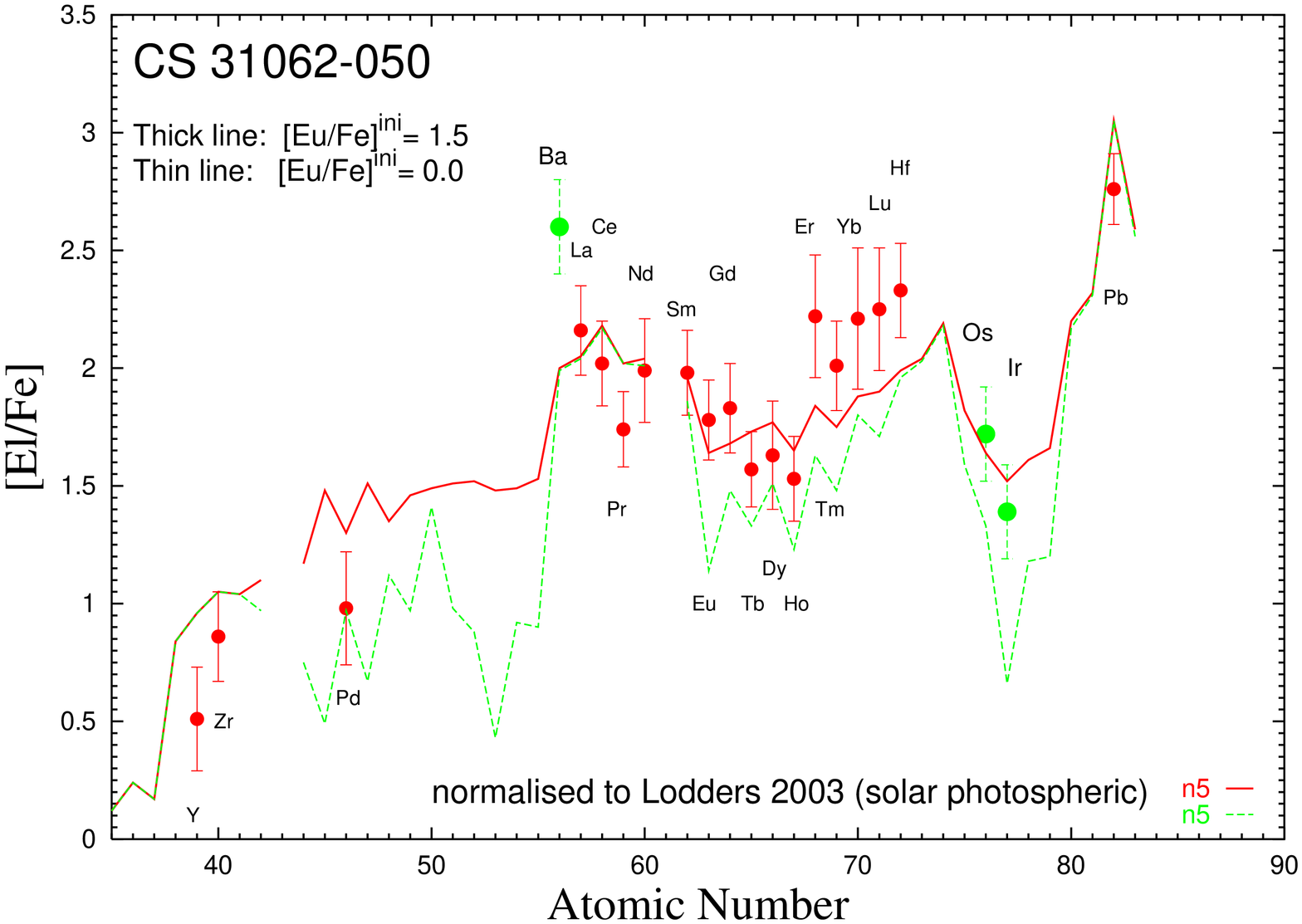}
\includegraphics[angle=-90, width=10cm]{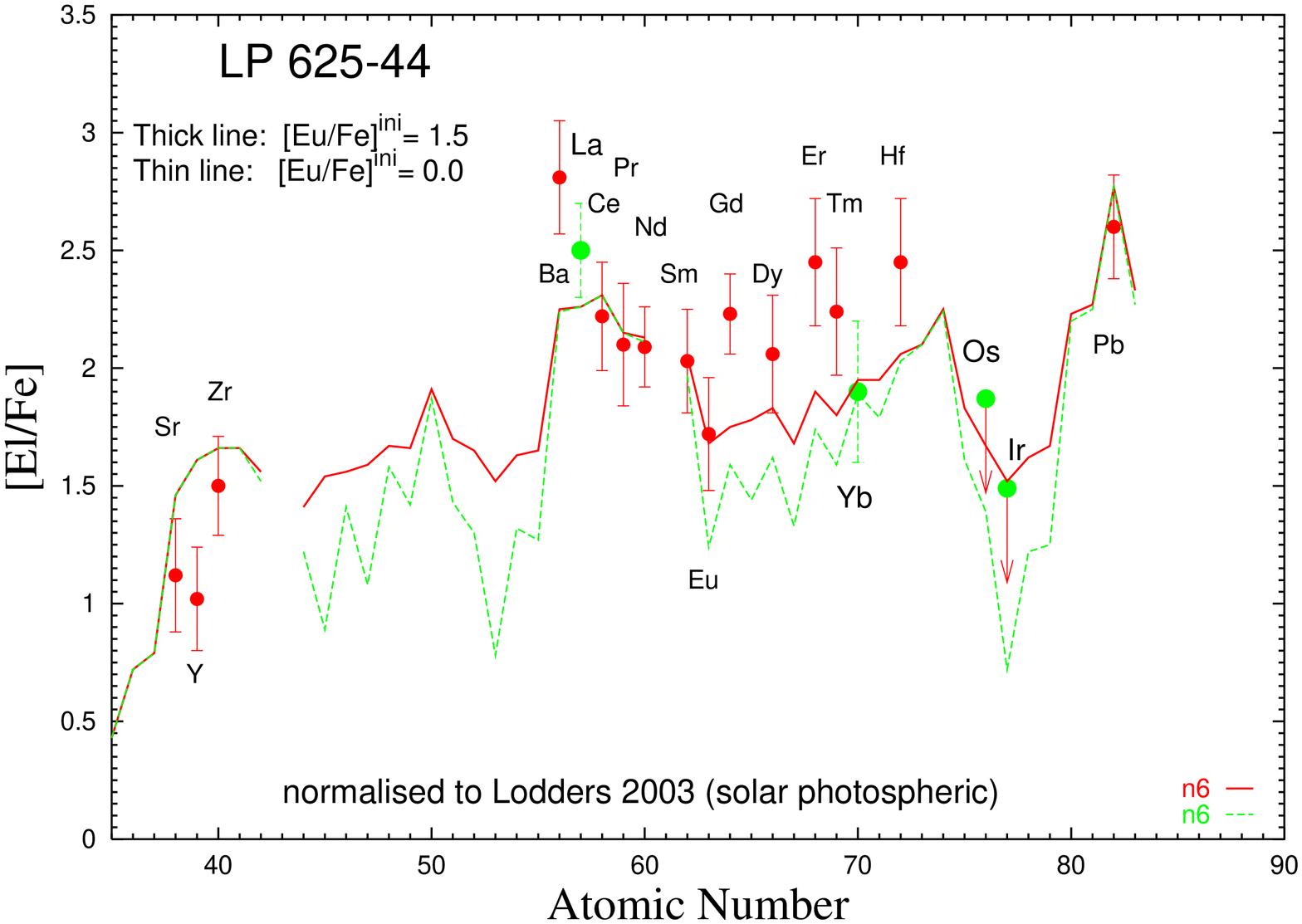}

\caption[]{Comparison of the calculated abundance patterns with the
observational results. The abundances are normalized to the solar
photospheric abundances provided by \citet{lodders03}. In the upper
panel, the Ba, Os, and Ir abundances of {\CS} measured in the present
work and those of other elements adopted from \citet{johnson04} are
shown along with the AGB s-process models for an initial mass of 1.3
M$_{\odot}$, [Fe/H]$=-2.42$, and the ST/10 case (a choice of $^{13}$C
concentration in the $^{13}$C pocket). The solid line indicates the
model assuming the initial enhancement of r-process elements
normalized to [Eu/Fe] $= +1.5$, while the dashed line is the model
predictions without such initial enhancements (see text for
details). The lower panel shows the same comparison, but for
{\LP}. The La and Yb abundances, and the upper limits of Os and Ir
abundances, determined by the present work are shown along with the
abundances of other elements adopted from \citet{aoki02a}. The models
assuming an initial mass of 1.3 M$_{\odot}$, [Fe/H]$=-2.70$, and ST/30
are shown in this diagram.}

\label{fig:comp}
\end{center}
\end{figure}

Figure \ref{fig:comp} shows a comparison of the observed abundance
patterns of heavy neutron-capture elements for {\CS} and {\LP} with
the FRANEC model calculations (Straniero et al. 2003; see also Zinner
et al. 2006), which are obtained for an AGB star of mass
1.3~M$_{\odot}$ with metallicity of [Fe/H] $=-2.4$ and [Fe/H] $=-2.7$,
respectively. The dotted lines indicate the results of the calculation
for which the abundance pattern of neutron-capture elements in
solar-system material, scaled to the model metallicity, is assumed for
the initial composition (here we refer to this as the standard
model). In the comparisons of the model calculations with
observational data, we give a priority to the abundances of the
elements near the three peaks of the s-process (Zr, La--Nd, and Pb; we
did not give a priority to Ba because of the difficulty mentioned
above). The abundance pattern of {\CS} for La--Sm and Pb agrees well
with the model prediction, while Eu, Gd, and Er--Hf exhibit excesses.
The Os and Ir abundances obtained by the present work are also clearly
higher than the predictions of the standard model.

The observed excesses of Os and Ir in {\CS}, as well as other elements such as
Eu, can be explained by assuming a large contribution to the
atmosphere of this star due to r-process nucleosynthesis. The
abundance pattern of heavy neutron-capture elements produced by the
(main) r-process is known to agree very well with that of the
r-process component in solar-system material
\citep[e.g. ][]{sneden03}. Hence, we assume excesses of the
neutron-capture elements with the solar-system r-process abundance
pattern as the initial abundances for the calculation. A good fit to
the observed abundances of Eu, Os, and Ir is found when the r-process
abundance pattern, normalized to [Eu/Fe]$ =+1.5$, is assumed for the
initial composition of this star. The abundance pattern obtained by
this calculation is shown by the solid line in Figure
\ref{fig:comp}. This assumption for the initial composition does not
significantly affect the final abundances of light neutron-capture
elements (Sr, Y, and Zr), nor the heavy s-process elements (Ba--Sm)
and Pb, because the contribution of the s-process is dominant. In
contrast, the abundances of Eu, Ir, Os, and several other heavy
neutron-capture elements (Gd--Yb), are significantly enhanced by this
assumption for the initial composition. We note that the Pd abundance
in {\CS} is better explained by the corresponding AGB model with no
initial r-process enrichment. However, the Pd abundance produced by
the r-process could be lower than the ``r-process component'' of the
solar-system Pd abundance. According to the abundance patterns found
in r-process-enhanced metal-poor stars
\citep[e.g.][]{hill02,sneden03}, the yields of the main r-process has
[Pd/Eu]$\sim -0.5$, while the r-process component in solar-system
material has [Pd/Eu]$=-0.25$. This suggests that the initial Pd
abundance assumed in our calculation is overestimated by about
0.2~dex.

%The Os and Ir abundances in {\CS} determined in the present work
%confirm the large contribution of the r-process to the abundance
%pattern of neutron-capture elements of this star. \citet{johnson04}
%claimed the requirement of a larger s-process contribution to Eu than
%that predicted by s-process models in order to account for the
%abundance pattern of this star. However, Figure \ref{fig:comp} shows
%that the overall abundance pattern of this object is well reproduced
%by our model, given the uncertainties of abundance measurements and
%model calculations. 

The overall abundance pattern of {\CS}, including Eu, Os and Ir, is
fit better by our model assuming excesses of r-process elements in the
initial abundances. The abundance pattern, in particular for Os and
Ir, cannot currently be explained without a large contribution of the
r-process to the neutron-capture elements in this star. Some elements
still show significant deviation from this model (e.g. Er--Hf),
indicating that our model assuming the excesses of neutron-capture
elements with the solar-system r-process abundance pattern is not a
final solution. However, the abundances of Os and Ir, which are almost
pure r-process elements, clearly request a large contribution of the
r-process or of some unknown process to {\CS} (see next section). For
{\LP}, we could only obtain upper limits on the Os and Ir abundances;
they are both lower than the measured abundances of these species for
{\CS}. From this result, we can only conclude that the contribution of
the r-process to the Os and Ir in {\LP}, if any, is smaller than that
in {\CS}. The higher [La/Eu] of this star ([La/Eu] $= +0.74$), as
compared to that of {\CS} ([La/Eu] $=+0.33$), also supports this
conclusion. It should be noted that the model predictions are scaled
to fit the three s-process-element abundance peaks, and the agreement
is not good for La--Hf.  However, the relatively low abundance ratio
of Pb with respect to the second s-process peak (e.g., Ba, La) in
{\LP}, as well as in several other objects, is a problem also found by
other recent observations \citep[e.g. ][]{aoki02b}, and further
s-process modeling for such objects is desired.

%The overall abundance pattern of {\CS}, including Eu, Os and Ir, is
%reproduced by our model assuming excesses of r-process elements as the
%initial abundances, though some elements show significant deviation
%from the model prediction (e.g. Er--Hf). This confirms the large
%contribution of the r-process to the abundance pattern of
%neutron-capture elements of this star (see next section).

%and
%the disagreement would reflect the current limitation of our
%understanding for the s-process, rather than the contribution of the
%r-process.

%However, the overall abundance pattern is still better
%explained by the model assuming initial excesses of the r-process
%elements, because of the large enhancements of Gd--Yb.

\section{Discussion and Concluding Remarks}

%Previous abundance studies have shown that {\CS} is a CEMP-r/s star,
%based on the observed excess of Eu compared to the abundance pattern
%produced by s-process nucleosynthesis at low metallicity estimated by
%recent AGB models. Our present measurements of Os and Ir abundances
%confirm the excess of r-process elements in this object. 

%Although a full understanding of the origin of such CEMP-r/s stars is beyond
%the scope of this Letter, 

There is no astrophysical model that can well explain the abundance
pattern of CEMP r/s stars. The model is required to explain the
following two observational facts: (1) More than ten CEMP-r/s stars
that exhibit very large excesses of Eu ([Eu/Fe]$>+1$) are already
known, while only a few r-process-enhanced stars without excesses of
s-process elements are known; (2) The CEMP-r/s stars exhibit larger
enhancements of heavy neutron-capture elements, on average, than the
stars having only s-process enhancements \citep[][]{jonsell06}. To
account for these constraints, a modified s-process model which yields
abundance patterns that are quite different from those of standard
models might be preferred, as has been discussed by previous authors
\citep[see ][ and references therein]{jonsell06}. However, the high Os
and Ir abundances, as well as the high Eu abundance, found in {\CS}
appear to exclude this possibility, because no s-process model known
to date predicts such high yields of these elements. We confirmed that
Os and Ir are not efficiently produced by our AGB models, even if
model parameters (initial mass, efficiency of mixing to produce the
neutron source $^{13}$C) are significantly changed. 
If a single astrophysical process capable of producing the observed
abundance patterns of r/s stars exists, it must be significantly
different than s-process production by AGB stars known so far.

%If a single process that can reproduce the abundance patterns of r/s
%stars should exist, the process should be significantly different from
%the s-process in AGB stars known so far.

As mentioned in the previous section, the abundance pattern of
neutron-capture elements in {\CS} is better accounted for by a model
that assumes an initial composition for the binary system with large
excesses of heavy elements having a scaled solar r-process abundance
pattern.  Thus, we argue that this system was formed from a parent
cloud polluted by a supernova that yielded r-process elements, which
might have also triggered the formation of next-generation low-mass
binary stars \citep{vanhala98}, while the s-process elements were
provided by the former primary star during its AGB phase. A possible
alternative scenario is to assume that pollution of {\CS} with
r-process-enhanced material took place after s-process
nucleosynthesis. This might be the case if the progenitor was an
8--10~M$_{\odot}$ star that underwent s-process nucleosynthesis during
its AGB phase, followed by the production of r-process elements during
its subsequent supernova explosion \citep{wanajo06}. However, it is
not yet clear how much of the s-process material may be mixed within
the envelope of these stars (Doherty 2006; Siess 2006). These two
scenarios are, unfortunately, not easily distinguished by the
elemental abundance pattern of the low-mass stars that are currently
observed, because the majority of the r-process nuclei for which
excesses are assumed for the initial composition are not subject to
the s-process. Hence, the r- and s-processes behave as almost
independent contributors to the final yields.

Finally, we point out that, although {\CS} and {\LP} have similarly
large excesses of carbon and neutron-capture elements, their elemental
abundance patterns exhibit non-negligible differences.  {\CS} exhibits
a large excess of Pb, as expected from s-process models
\citep{aoki02b,johnson04}, while {\LP} has a smaller enhancement of Pb
\citep{aoki00,aoki02a}. The [La/Eu] ratios show some difference, as
mentioned in \S 3. In addition, the sodium (Na) abundance is
significantly different between the two stars
\citep{aoki02a,aoki06}. These differences suggest that the process(es)
responsible for the production of these elements may not be the same
in these two stars. More detailed comparisons of the elemental
abundances for these two stars are strongly desired.

%These differences suggest that the process responsible for
%the large excesses of neutron-capture elements, as well as
%carbon and nitrogen, in these two stars are not same. 

\acknowledgments

W.A. is grateful for useful discussions with Drs. N. Iwamoto,
T. Kajino, and G. J. Mathews. R.G. acknowledges support by the Italian
MIUR-FIRB Project "Astrophysical Origin of Heavy Elements beyond Fe"
T.C.B. acknowledges partial support from a series of grants awarded by
the US National Science Foundation, most recently, AST 04-06784, as
well as from grant PHY 02-16783; Physics Frontier Center/Joint
Institute for Nuclear Astrophysics (JINA). J.E.N. acknowledges support
from Australian Research Council grant DP0342613.

\clearpage
\begin{figure} 
\includegraphics[width=14cm]{f1.ps}
\caption[]{Comparisons of synthetic spectra for the \ion{Ir}{1}
$\lambda$3513~{\AA} and the \ion{Os}{1} $\lambda$4260~{\AA} features
with the observed ones in {\CS} and {\LP}. The assumed abundances are
$\log \epsilon$(Os)$=0.75 \pm 0.20$ and $\log \epsilon$(Ir)$=0.35\pm
0.20$ for {\CS}, while those for {\LP} are $\log \epsilon$(Os)$=0.60$
(adopted upper limit) and 0.25 (for a comparison), and $\log
\epsilon$(Ir)$=0.15$ (adopted upper limit) and $-0.05$ (for a
comparison). The dotted lines indicate the spectra calculated assuming
no Os or Ir.}
\label{fig:iros}
\end{figure} 

\begin{deluxetable}{lccccccccc}
\tablewidth{0pt}
\tablecaption{ATMOSPHERIC PARAMETERS AND ABUNDANCES RESULTS\label{tab:res}}
\startdata
%\noalign{smallskip}
\tableline
%\hline
\tableline
                           & \multicolumn{4}{c}{\CS}    & & \multicolumn{4}{c}{\LP}  \\
\cline{2-5}\cline{7-10}
$T_{\rm eff}$(K)           & \multicolumn{4}{c}{5500}   & & \multicolumn{4}{c}{5500}  \\
$\log g/$[cm~s$^{-2}$]     & \multicolumn{4}{c}{2.7}    & & \multicolumn{4}{c}{2.5} \\
$[$Fe/H]                   & \multicolumn{4}{c}{$-2.3$} & & \multicolumn{4}{c}{$-2.7$} \\
$v_{\rm turb}$ (km~s$^{-1}$) & \multicolumn{4}{c}{1.3}    & & \multicolumn{4}{c}{1.2} \\
\cline{2-5}\cline{7-10}
    & $\log \epsilon$ & [X/Fe] & $\sigma$ & $\log \epsilon_{\rm JB04}$\tablenotemark{a}  & &  $\log \epsilon$ & [X/Fe] & $\sigma$ & $\log \epsilon_{\rm A02}$\tablenotemark{a} \\ 
\tableline
%La & &&&&&&\\ 
Ba & 2.35    & +2.60 & 0.2  & 2.61 & & 2.31    & +2.86 & 0.2  & 2.31 \\
La & 0.95    & +2.24 & 0.06 & 0.93 & & 0.91     & +2.50 & 0.07  & 0.90 \\
Eu & 0.01    & +1.91 & 0.05 & $-0.07$ & & $-0.44$ & +1.76 & 0.05 & $-0.48$ \\
Yb & 0.5:    & +1.8: & 0.3  & 0.76 & & 0.3:    & +1.9: & 0.3 & \nodata\\
Hf & 0.67    & +2.21 & 0.1  & 0.67 & & 0.55    & +2.39 & 0.1  & 0.48 \\
Os & 0.75    & +1.72 & 0.2  & \nodata& & $<0.60$ & $<1.87$ & \nodata & \nodata\\
Ir & 0.35    & +1.39 & 0.2  & \nodata& & $<0.15$ & $<1.49$ & \nodata & \nodata\\
Pb & 2.45    & +2.87 & 0.1  & 2.46 & & 1.95    & +2.67    & 0.1  & 1.90 \\ 
\tableline
\enddata
\tablenotetext{a}{Results obtained by \citet{johnson04} and \citet{aoki02a}.}
\end{deluxetable}

\end{document}